%% file: main.tex
\documentclass{mva_style}
\usepackage{graphicx}
\usepackage{color}
\usepackage{tabularx}
\usepackage{booktabs}
\usepackage{multirow}
\usepackage{amsmath}
\usepackage{textcomp}
\usepackage{amsmath}
\usepackage{amssymb}
\usepackage{url}
\pdfoutput=1
\usepackage[
    colorlinks=true,
    linkcolor=black,      
    citecolor=black,      
    colorlinks
]{hyperref}

\usepackage{caption}      
\usepackage[table,xcdraw]{xcolor}
\usepackage{colortbl}

\definecolor{midcolor}{RGB}{253, 243, 209}   
\definecolor{downcolor}{RGB}{219, 229, 214}  
\finalcopy 
\definecolor{yzybest}{rgb}{0.98, 0.8, 0.8}    
\definecolor{yzysecond}{rgb}{0.99, 0.88, 0.77} 
\definecolor{yzythird}{rgb}{1.0, 1.0, 0.8}     

\newcommand{\bestcolor}{\cellcolor{yzybest}}
\newcommand{\secondcolor}{\cellcolor{yzysecond}}

\begin{document}
\title{ZECO: ZeroFusion Guided 3D MRI Conditional Generation}

\author{
  Feiran Wang$^1$, Bin Duan$^2$, Jiachen Tao$^3$, Nikhil Sharma$^1$, Dawen Cai$^2$, Yan Yan$^{3, \dagger}$ \\
  $^1$Illinois Institute of Technology, $^2$University of Michigan, $^3$University of Illinois Chicago \\
}

\maketitle
\begingroup
\renewcommand{\thefootnote}{}
\footnotetext[1]{Project page: \url{https://brack-wang.github.io/ZECO_web/}}
\footnotetext[2]{ $^\dagger$Corresponding Author}
\endgroup

\input{sec/0_abstract}

\input{sec/1_introduction}

\input{sec/2_relatedwork}

\input{sec/3_methods}

\input{sec/4_experiments}

\input{sec/5_conclusion}

\bibliographystyle{ieee}
\bibliography{main}






\end{document}

%% file: sec/0_abstract.tex
\section*{\centering Abstract}
\textit{
Medical image segmentation is crucial for enhancing diagnostic accuracy and treatment planning in Magnetic Resonance Imaging (MRI). However, acquiring precise lesion masks for segmentation model training demands specialized expertise and significant time investment, leading to a small dataset scale in clinical practice. In this paper, we present \textbf{ZECO}, a ZeroFusion guided 3D MRI conditional generation framework that extracts, compresses, and generates high-fidelity MRI images with corresponding 3D segmentation masks to mitigate data scarcity. To effectively capture inter-slice relationships within volumes, we introduce a Spatial Transformation Module that encodes MRI images into a compact latent space for the diffusion process. Moving beyond unconditional generation, our novel ZeroFusion method progressively maps 3D masks to MRI images in latent space, enabling robust training on limited datasets while avoiding overfitting. ZECO outperforms state-of-the-art models in both quantitative and qualitative evaluations on Brain MRI datasets across various modalities, showcasing its exceptional capability in synthesizing high-quality MRI images conditioned on segmentation masks.
}

%% file: sec/1_introduction.tex
\section{Introduction}
\vspace{-0.18cm}
\label{sec:intro}
Medical image segmentation~\cite{ma2024segment, wolleb2022diffusion, rahman2023ambiguous, wu2024medsegdiff} is indispensable for facilitating accurate diagnosis in clinical practice, and has achieved promising progress in segmenting lesion masks for 2D medical images. However, accurately annotated 3D MRI images are scarce, leading to limited dataset sizes and increased risk of model overfitting. The limitation stems from two key challenges: (1) acquiring MRI data in clinical environments is constrained by high scanning costs and safety concerns~\cite{kim2024adaptive}; and (2) annotating 3D volumes is exceptionally time-consuming and requires substantial expertise from medical specialists. To overcome the dataset scarcity, image generation methods offer a promising solution for synthesizing large quantities of realistic medical images based on the characteristics and distribution of limited sample data.

\input{images/teaser}
Recent advances in image generation~\cite{kazeminia2020gans, armanious2020medgan, goodfellow2014generative, lyu2022conversion, rombach2022high, sohl2015deep}  have laid the groundwork for generating diverse, high-resolution medical images. Among these advances, diffusion models~\cite{gungor2023adaptive, jiang2023cola, lyu2022conversion, rombach2022high, sohl2015deep} have emerged as the dominant approach, offering stable training processes and enhanced image fidelity.
Med-DDPM~\cite{dorjsembe2023conditional} generates 3D brain MRIs by incorporating conditions into each step of the diffusion process, while SegGuidedDiff ~\cite{konz2024anatomically} achieved controllable medical image generation by processing 3D images as 2D slice stacks. However, two key challenges remain: (1) Directly fine-tuning a limited set of conditions data on a  pretrained diffusion model risks overfitting and catastrophic forgetting~\cite{hu2021lora, ruiz2023dreambooth, zhang2023adding}. (2) The 3D nature of medical images requires models to capture inter-slice relationships to generate consistent and coherent 3D images~\cite{kim2024adaptive}.

To address these challenges, we propose \textbf{ZECO}, the \textbf{Ze}roFusion Guided 3D MRI \textbf{Co}nditional Generation to synthesize high-quality MRI images conditioned on segmentation masks. To capture the anatomical relationships within the 3D volume, we first introduce the Spatial Transformation Module to encode MRI into the latent space, enabling efficient training and preserving complex inter-slice correlations. 
We then integrate a diffusion model with 3D-UNet to synthesize MRI images in the latent space without conditions. To avoid directly training conditions on pretrained diffusion models with the risk of overfitting, we introduce a ZeroFusion structure that progressively learns the relationships between MRI and their embedded conditions.

We compared ZECO with baselines and state-of-the-art (SOTA) works on the BraTS 2016 and BraTS 2020 MRI datasets, demonstrating superior performance both qualitatively and quantitatively. As shown in Figure \ref{fig:result}, ZECO synthesizes realistic yet distinct MRI slices based on segmentation masks for both FLAIR and T1 modalities.  
Our key contributions are summarized as follows:
i) We introduce a ZeroFusion Structure, a conditional medical generative framework designed to train on limited mask-specific conditions; ii) We integrate a Spatial Transformation Module to capture the spatial relationships within 3D medical images; iii) Extensive experimental results and ablation studies demonstrate the efficacy of our model.

%% file: images/teaser.tex
\setlength{\textfloatsep}{5pt}
\begin{figure}[!t]
  \centering
  \includegraphics[width=\linewidth]{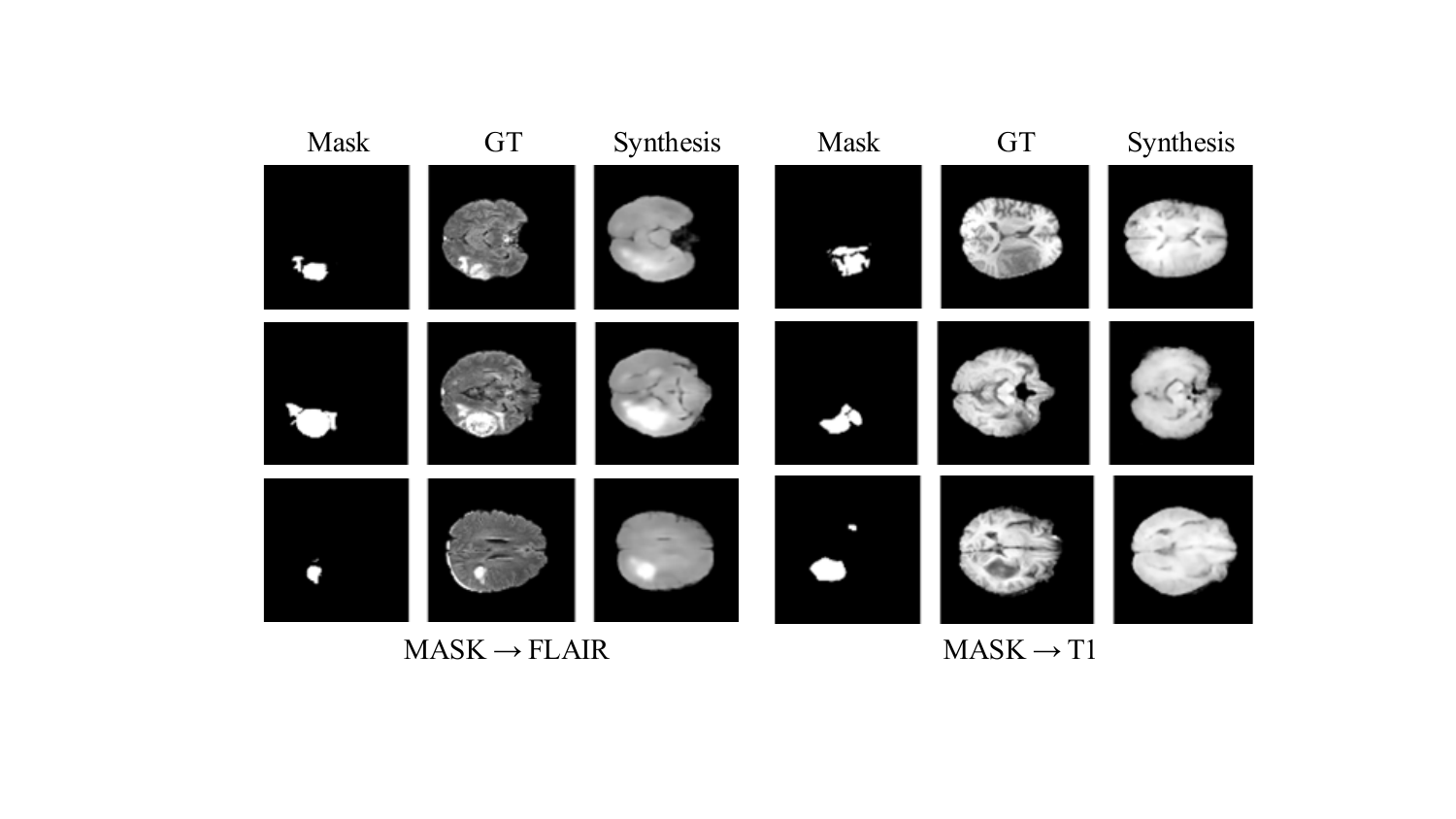}  
  \caption{ZECO generates FLAIR (left) and T1 (right) MRI modalities conditioning on segmentation masks.}
  \label{fig:result}
\end{figure}

%% file: sec/2_relatedwork.tex
\section{Related Work}
\vspace{-0.18cm}
\label{sec:related_work}
Generative model~\cite{jiang2023cola, rombach2022high, yurt2022progressively,kazeminia2020gans, armanious2020medgan, goodfellow2014generative} has achieved great success on the medical images.
A widely adopted approach for medical image generation is based on Generative Adversarial Networks (GANs)~\cite{skandarani2023gans, kazeminia2020gans, armanious2020medgan, goodfellow2014generative}. GANs use a generator to create data and a discriminator to evaluate its authenticity, with both networks competing with each other to improve the realism of the generated data. 
Although GANs can generate high-quality images, they suffer from training instability that leads to convergence failures~\cite{wiatrak2019stabilizing}. Additionally, they also experience mode collapse, producing limited variations in output~\cite{nie2018medical}. Therefore, we need more robust and stable approaches that can consistently produce diverse and realistic medical images.

Denoising Diffusion Probabilistic Models (DDPMs)~\cite{ho2020denoising, zhuang2023semantic, ronneberger2015u, rombach2022high, yurt2022progressively} have emerged as a powerful generative approach by gradually adding noise to training data and then learning to reverse this process step-by-step to reconstruct the original data.
They can stably generate realistic and high-resolution data and have shown significant advancements in medical image generation, such as MRI synthesis~\cite{jiang2023cola, rombach2022high, yurt2022progressively}, 
and especially conditional generation~\cite{khader2023denoising, pinaya2022brain, dorjsembe2023conditional, konz2024anatomically}. Med-DDPM~\cite{dorjsembe2023conditional}  introduced a latent diffusion model combined with a GAN to encode 3D images for efficient training, achieving high-fidelity unconditional generation. SegGuidedDiff~\cite{konz2024anatomically}  is designed for conditional generation using a random mask ablation strategy but fails to capture inter-slice relationships. Our method leverages the efficiency of the latent diffusion model while enhancing conditional generation by preserving spatial relationships.

%% file: sec/3_methods.tex
\section{Methods}
\vspace{-0.18cm}
\label{sec:pagestyle}

\input{images/pipeline}

In this section, we introduce \textbf{ZECO}, a conditional medical generative model designed to produce MRI images precisely guided by 3D segmentation masks, consisting of a  Spatial Transformation Module, the 3D Diffusion Process, and the ZeroFusion Structure.

\subsection{Spatial Transformation Module}
\label{STM}
To efficiently process and generate high-resolution 3D images, we introduce the Spatial Transformation Module \textbf{(STM)} originated from GANs, which transforms 3D images into a latent space  $Z$ and reconstructs images from the latent space. 
To process the 3D MRI, we elevate from 2D convolutions to 3D convolutions to suit the 3D volume inputs. To further reduce computational cost and focus on local features, we employ a patch discriminator that divides images into smaller patches and performs predictions at the patch level. 
The encoder compresses MRI into a compact latent representation. During quantization, continuous latent vectors are mapped to their nearest discrete counterparts in the codebook. 
Finally, the decoder reconstructs MRI from the latent space, shown in Figure \ref{fig:architecture} (a).
The total loss($\mathcal{L}_{\text{stm}}$) combines multiple objectives: the reconstruction loss($\mathcal{L}_{\text{r}}$) ensures image fidelity, adversarial loss ($\mathcal{L}_{\text{a}}$) promotes realism, perceptual loss($\mathcal{L}_{\text{p}}$) aligns high-level features, codebook loss($\mathcal{L}_{\text{cb}}$) maintains the closeness of the latent vectors to the codebook, and commitment loss($\mathcal{L}_{\text{cm}}$) encourages the encoder to align with codebook entries.
\begin{equation}
  \mathcal{L}_{\text{stm}} = \mathcal{L}_{\text{r}} + \lambda_{\text{a}} \mathcal{L}_{\text{a}} + \lambda_{\text{p}} \mathcal{L}_{\text{p}} + \lambda_{\text{cb}} \mathcal{L}_{\text{cb}} + \lambda_{\text{cm}} \mathcal{L}_{\text{cm}}
\end{equation}
where $\lambda$ values control the weight of each loss term.

\subsection{3D Diffusion Process}
We implement a straightforward yet effective 3D diffusion model, generating unconditional MRI images within the latent space encoded through STM.  As shown in Figure \ref{fig:architecture} (b), our diffusion model is built on a 3D U-Net architecture with four encoder and decoder blocks, each consisting of three ResNet blocks and one attention block. The middle block, positioned at the bottleneck, consists of two ResNet blocks and an attention block to refine latent features. 
During the reverse denoising process, the 3D U-Net is trained for unconditional generation, transforming the input feature map $z_t$ into the output feature map $z_{t-1}$ at timestep t - 1.

 In latent space, feature images at time \( t \) are represented as \( z_t \in \mathbb{R}^{c \times h \times w \times d} \), where \( c \), \( h \), \( w \), and \( d \) denote the channels, height, width, and depth, respectively. The forward process, \( q(z_t|z_{t-1}) \), progressively adds noise to the input latent data \( z_0 \) until it becomes nearly Gaussian noise \( z_T \). The reverse process learns to remove the noise and recover the input data distribution, which can be described as sampling from the distribution \( p(z_{t-1}|z_t) \) by predicting the added noise:
\begin{equation}
    \mu_\theta = \frac{1}{\sqrt{\alpha_t}} \left(z_t - \frac{\beta_t}{\sqrt{1 - \bar{\alpha}_t}} \epsilon_\theta(z_t, t)\right)
\end{equation}
where \( \epsilon_\theta(z_t, t) \) represents the predicted noise, \( \alpha_t \) and \( \beta_t \) are derived from the noise schedule. Specifically, \( \alpha_t = 1 - \beta_t \), where \( \beta_t \) controls the variance of the noise added at each timestep \( t \). \( \bar{\alpha}_t \) represents the cumulative product of \( \alpha_t \) up to timestep. To maintain mathematical consistency with the forward process, which introduces Gaussian noise at each step, the reverse process models the feature map at timestep t-1 as follows based on the predicted noise $\mu_\theta(z_t, t)$:

\begin{equation}
z_{t-1} = \mathcal{N}(z_{t-1}; \mu_\theta(z_t, t), \Sigma_{\theta}(z_t, t))
\end{equation}

\subsection{ZeroFusion Structure}
We propose ZeroFusion Structure, a novel architecture designed to enhance pretrained diffusion models for MRI image generation under 3D segmentation mask guidance, as illustrated in Figure \ref{fig:architecture} (c). To keep the model's capacity for synthesizing high-quality MRI images while avoiding overfitting, we freeze the weights of the 3D diffusion model and create a trainable copy as ZeroFusion. During the training process, empowered with the ability to generate MRI unconditionally, ZeroFusion is conditioned on the segmentation masks with input MRI volumes encoded in the latent space. This approach implicitly maps masks to images while learning the spatial relationships within 3D volumes.

To gradually adjust ZeroFusion without destabilizing the model's generative ability, we introduce 3D zero modules $\mathcal{H}(\cdot)$, a series of $1 \times 1 \times 1$ convolutional layers. We strategically integrated 3D Zero modules into ZeroFusion's middle sample and down sample process, generating the middle-level feature $f_m$ and down-level feature $f_d$ respectively. These features contribute to reconstructing the latent image from \( z_{t} \) to \( z_t-1 \) combined with features $\mu_\theta$ from the frozen model. 
To ensure that no random noise is added, we initialize the weights and biases of the 3D zero modules to zero, guaranteeing that the trainable copy initially behaves identically to the pretrained model. The complete ZeroFusion computes the noise $\mu_z(t)$ at timestep $t$ as following:

\begin{equation}
\mu_z = \mu_\theta(z_{t}, t) + \mathcal{H}(f_d(z_{t};c) + \mathcal{H}(f_m(z_{t};c))) 
\end{equation}
where $c$ represents the conditioning vector in latent space. Given the conditions at time step $t$ and a segmentation mask, ZeroFusion estimates the noise that was added to the noisy image. With its foundation in a pretrained model, ZeroFusion can rapidly synthesize MRI morphology. The flexibility provided by the 3D zero modules enables ZeroFusion to effectively map segmentation masks to corresponding MRI images. The loss objective $\mathcal{L}$ is defined as:
\begin{equation}
    \mathcal{L} = \mathbb{E}_{z_0, t \in [0, T], c, \epsilon \sim \mathcal{N}(0, 1)} \left[ \left\| z_t - \mu_z (z_{t}, t, c) \right\|_2^2 \right]
\end{equation}

%% file: images/pipeline.tex
\setlength{\textfloatsep}{15pt}
\begin{figure*}[t]  
    \centering         
    \includegraphics[width=0.8\linewidth]{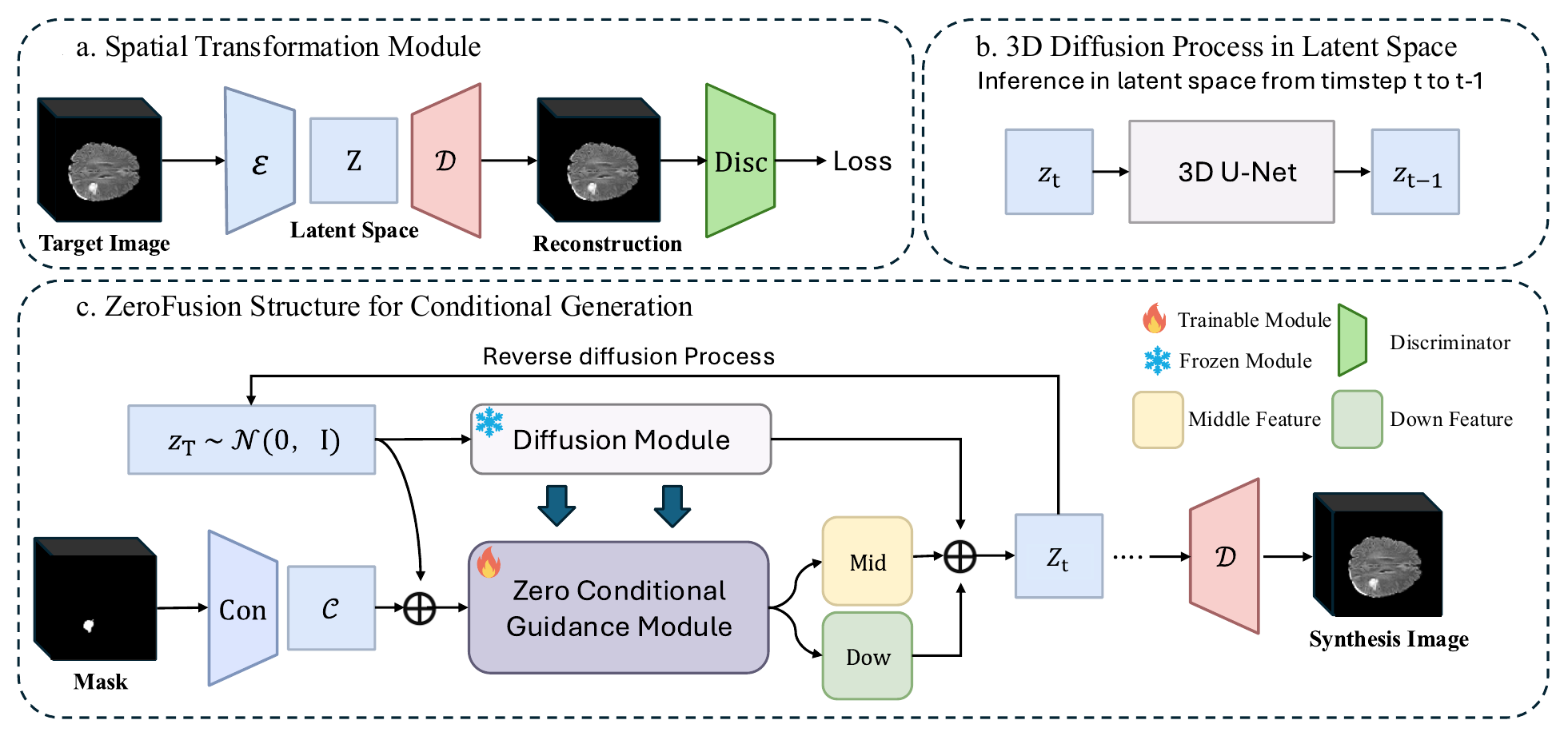}  
    \caption{Pipeline. (a) The Spatial Transformation Module encodes MRI images into latent space. (b) The 3D U-Net reconstruct latent images across timesteps in the reverse diffusion process. (c) The ZeroFusion generates \colorbox{midcolor}{Middle-feature} and \colorbox{downcolor}{Down-feature} conditioned on segmentation masks $\mathcal{C}$ for controllable generation. }
    \label{fig:architecture}
\end{figure*}

%% file: sec/4_experiments.tex
\section{Experiment Settings}
\vspace{-0.18cm}
\label{sec:majhead}
\noindent\textbf{Datasets.}
We evaluated our model on the BraTS 2016~\cite{69752101} and BraTS 2020~\cite{mehta2021qu} datasets. While both datasets are widely used for brain tumor segmentation, BraTS 2020 offers a larger and more challenging set of tumor cases compared to BraTS 2016. We train our model using the FLAIR and T1 modalities, as they provide complementary information: FLAIR highlights areas of abnormal fluid, making it suitable for detecting tumor-associated edema, while T1 offers high-resolution anatomical details, essential for identifying precise tumor boundaries.

\noindent\textbf{Metrics.}
Performance was evaluated using four key metrics: SSIM~\cite{ssim}, MS-SSIM~\cite{wang2003msssim}, MMD~\cite{gretton2012mmd}, and PSNR~\cite{hore2010psnr}. SSIM and MS-SSIM assess image similarity, focusing on structural and textural details. For 3D images, they were applied to individual slices and aggregated. MMD quantifies distributional differences between generated images and ground truth, while PSNR evaluates reconstruction fidelity.

\input{table/compare_all}

\noindent\textbf{Implementation Details.}
We trained ZECO on an NVIDIA RTX 6000. The Adam optimizer had a learning rate of $1 \times 10^{-4}$ for the 3D diffusion model, and $2.5 \times 10^{-5}$ for the ZeroFusion Structure. All modules were trained from scratch with a batch size of 6.

\input{images/compare}

\section{Experiment Results}
\vspace{-0.18cm}
\noindent\textbf{Quantitative Results.} 
Our method demonstrates superior performance compared to existing generative models, as shown in Table \ref{tab:3DMCG_results}. We evaluated ZeroFusion against the baseline of Latent Diffusion Model (LDM) ~\cite{rombach2022high}, which we modified for conditional generation by concatenating segmentation masks with noise during the diffusion process. Our model significantly outperforms Med-DDPM, the current state-of-the-art diffusion-based method that incorporates conditioning signals directly into the denoising steps. While SegGuidedDiff can synthesize high-fidelity images for slices with conditional masks, it produces anatomically implausible results for slices without masks in MRI volumes due to insufficient spatial relationship modeling. This limitation highlights the importance of our approach's ability to maintain spatial coherence throughout the volume. Specifically, ZECO achieves substantial improvements across all evaluation metrics consistently across experimental settings.

\noindent\textbf{Qualitative Results.}
Our method excels at generating 3D MRI volumes of FLAIR modality compared to state-of-the-art models. As illustrated in Figure \ref{fig:compare}, when processing sequences of segmentation masks, ZECO consistently generates precise and anatomically coherent slices across the entire volume. In contrast, SegGuidedDiff produces inconsistent results in slices without explicit conditioning masks (Row 1), leading to anatomical discontinuities and consequently suboptimal quantitative performance. 
The improved spatial coherence is particularly evident in regions where segmentation masks provide limited guidance, demonstrating our model's robust ability to maintain anatomical fidelity throughout the entire 3D structure.

\input{table/ablation}

\noindent\textbf{Ablation Study of ZeroFusion}
We evaluate ZeroFusion's impact on model performance in Table \ref{tab:ablation}. Removing ZeroFusion decreases SSIM and MS-SSIM scores across all modalities, while incorporating it improves PSNR, produces sharper generations, and reduces MMD, indicating better alignment with real images. ZeroFusion increases SSIM by 1.8$\%$ and 1.1$\%$ for FLAIR modality on BraTS 2016 and 2020 datasets respectively, with improvements of 2.5$\%$ and 1.7$\%$ for T1 modality. These consistent enhancements demonstrate ZeroFusion's effectiveness in learning implicit relationships between segmentation masks and MRI images while preserving the pretrained model's capabilities.

%% file: table/compare_all.tex
\begin{table*}[!t]
\vspace{-0.2mm}
\centering
\caption{Quantitative evaluation across datasets, with \colorbox{yzybest}{best} and \colorbox{yzysecond}{second best} results highlighted for all metrics.}
\label{tab:3DMCG_results}
\resizebox{\textwidth}{!}{%
\begin{tabular}{llcccccccc}
\toprule
\multirow{2}{*}{Dataset} & \multirow{2}{*}{Method} & \multicolumn{4}{c}{MASK $\rightarrow$ FLAIR} & \multicolumn{4}{c}{MASK $\rightarrow$ T1} \\ 
\cmidrule(lr){3-6} \cmidrule(lr){7-10}
                         &                         & SSIM ↑        & MS-SSIM ↑     & MMD ↓         & PSNR ↑        & SSIM ↑        & MS-SSIM ↑     & MMD ↓         & PSNR ↑        \\ 
\midrule
\multirow{4}{*}{BraTS 2016} 
                         & LDM                    & 0.691 ± 0.035  & 0.675 ± 0.045  & 0.028 ± 0.047  & 21.202 ± 3.472 & 0.739 ± 0.021  & 0.732 ± 0.027  & 0.020 ± 0.028  & 22.048 ± 2.964 \\
                         & SegGuidedDiff          & 0.698 ± 0.068  & 0.711 ± 0.011  & 0.049 ± 0.066  & \secondcolor 22.056 ± 3.215 & 0.687 ± 0.018  & 0.720 ± 0.049  & 0.042 ± 0.026  & 22.056 ± 2.785 \\
                         & Med-DDPM               & \secondcolor 0.731 ± 0.037  & \bestcolor 0.856 ± 0.041  & \secondcolor 0.018 ± 0.054  &  21.901 ± 3.333 & \secondcolor 0.800 ± 0.021  & \secondcolor 0.847 ± 0.027  & \secondcolor 0.011 ± 0.039  & \secondcolor 23.144 ± 3.446 \\
                         & \textbf{Ours}          & \bestcolor 0.794 ± 0.033  & \secondcolor 0.837 ± 0.022  & \bestcolor 0.012 ± 0.026  & \bestcolor 23.218 ± 3.159 & \bestcolor 0.831 ± 0.027  & \bestcolor 0.875 ± 0.038  & \bestcolor 0.009 ± 0.039  & \bestcolor 24.491 ± 5.746 \\
\midrule

\multirow{4}{*}{BraTS 2020} 
                         & LDM                    & 0.796 ± 0.022  & 0.584 ± 0.027  & 0.054 ± 0.028  & 13.209 ± 2.964 & 0.741 ± 0.058  & 0.693 ± 0.046  & 0.025 ± 0.045  & 16.050 ± 2.223 \\
                         & SegGuidedDiff          & 0.759 ± 0.016  & 0.705 ± 0.019  & 0.052 ± 0.027  & \secondcolor 23.725 ± 8.120 & 0.623 ± 0.038  & 0.634 ± 0.047  & 0.069 ± 0.043  & 21.056 ± 1.845 \\
                         & Med-DDPM               & \secondcolor 0.835 ± 0.023  & \secondcolor 0.842 ± 0.096  & \secondcolor 0.018 ± 0.042  &  22.673 ± 7.008 & \secondcolor 0.849 ± 0.018  & \secondcolor 0.877 ± 0.083  & \bestcolor 0.013 ± 0.023  & \secondcolor 21.607 ± 4.235 \\
                         & \textbf{Ours}          & \bestcolor 0.851 ± 0.055  & \bestcolor 0.886 ± 0.013  & \bestcolor 0.012 ± 0.036  & \bestcolor 24.049 ± 8.695 & \bestcolor 0.858 ± 0.035  & \bestcolor 0.896 ± 0.011  & \secondcolor0.014 ± 0.031  & \bestcolor 22.697 ± 6.289 \\
\bottomrule

\end{tabular}%
}
\vspace{-2.0mm}
\end{table*}

%% file: images/compare.tex
\setlength{\textfloatsep}{5pt}
\begin{figure}[!t]
  \centering
  \includegraphics[width= \linewidth]{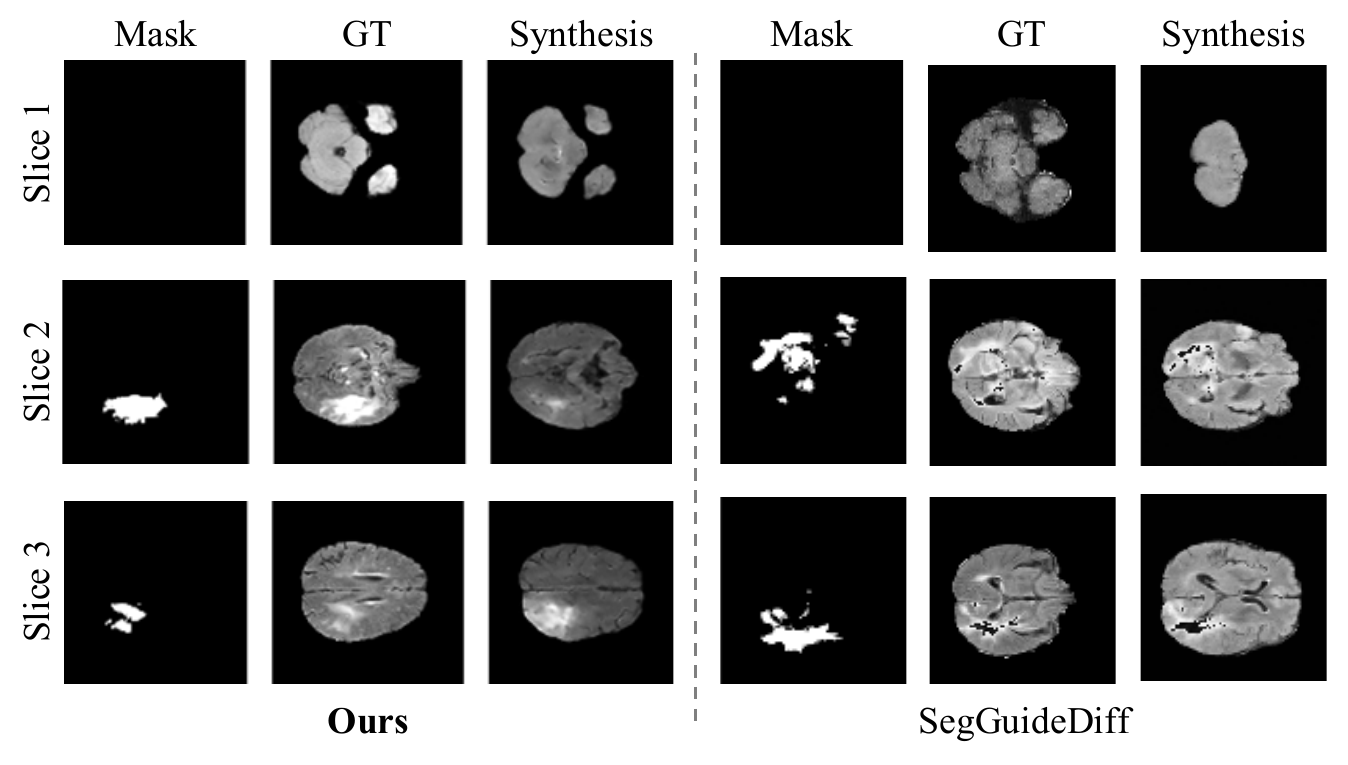}
  \vspace{-20pt}
  \caption{ ZECO generates coherent MRI slices, while SegGuidedDiff fails in regions without masks.
  }
  \label{fig:compare}
\end{figure}

%% file: table/ablation.tex
\setlength{\textfloatsep}{12pt}
\begin{table}[!t]
\centering
\caption{Performance comparison with and without ZeroFusion Structure across datasets and modalities.}
\vspace{-0.05cm}
\label{tab:ablation}
\renewcommand{\arraystretch}{1.5}  
\setlength{\tabcolsep}{4pt}  
\resizebox{\columnwidth}{!}{%
\begin{tabular}{c|>{\centering\arraybackslash}m{1.5cm}|c|c|c|c|c}
\hline
\textbf{Dataset} & \textbf{Modality} & \textbf{w/o} & \textbf{SSIM} ↑  & \textbf{MS-SSIM} ↑ & \textbf{MMD} ↓  & \textbf{PSNR} ↑ \\ \hline
\multirow{4}{*}{\shortstack{BraTS \\ 2016}} 
                    & \centering Flair  & -       & 0.776 ± 0.014    & 0.802 ± 0.026    & 0.013 ± 0.032    & 23.044 ± 5.352  \\ 
                    & \centering        & \checkmark & \textbf{0.794 ± 0.033} & \textbf{0.849 ± 0.021} & \textbf{0.012 ± 0.026} & \textbf{23.218 ± 3.159} \\ \cline{2-7}
                    & \centering T1     & -       & 0.806 ± 0.022    & 0.853 ± 0.029    & 0.010 ± 0.036    & 23.543 ± 4.532  \\ 
                    & \centering        & \checkmark & \textbf{0.831 ± 0.027} & \textbf{0.875 ± 0.038} & \textbf{0.008 ± 0.039} & \textbf{24.491 ± 5.746} \\ \hline
\multirow{4}{*}{\shortstack{BraTS \\ 2020}} 
                    & \centering Flair  & -       & 0.840 ± 0.019    & 0.843 ± 0.010    & 0.019 ± 0.064    & 21.835 ± 6.868  \\ 
                    & \centering        & \checkmark & \textbf{0.851 ± 0.055} & \textbf{0.886 ± 0.013} & \textbf{0.012 ± 0.036} & \textbf{24.049 ± 8.695} \\ \cline{2-7}
                    & \centering T1     & -       & 0.841 ± 0.032    & 0.829 ± 0.023    & 0.026 ± 0.048    & 19.014 ± 4.341  \\ 
                    & \centering        & \checkmark & \textbf{0.858 ± 0.035} & \textbf{0.896 ± 0.011} & \textbf{0.012 ± 0.026} & \textbf{22.697 ± 6.289} \\ \hline
\end{tabular}%
}
\end{table}

%% file: sec/5_conclusion.tex
\section{Conclusion}
\vspace{-0.18cm}
\label{sssec:subsubhead}
In this paper, we present \textbf{ZECO}, a ZeroFusion Guided 3D MRI Conditional Generation framework utilizing segmentation masks. 
Our primary contribution involves the introduction of the ZeroFusion structure that effectively generates medical images under customized conditions with unprecedented fidelity. Furthermore, our proposed Spatial Transformation Module significantly enhances the extraction of spatial interrelationships within complex 3D medical images. Experimental results demonstrate ZECO's superiority over existing SOTA methods across multiple modalities and datasets, providing a robust foundation for generating high-quality MRI images at scale.
The framework shows significant potential for extension to other medical imaging domains characterized by sparse labels, addressing data scarcity challenges and advancing diagnostic capabilities in clinical settings.

%% file: main.bbl
\begin{thebibliography}{10}\itemsep=-1pt

\bibitem{armanious2020medgan}
K.~Armanious, C.~Jiang, M.~Fischer, T.~K{\"u}stner, T.~Hepp, K.~Nikolaou, S.~Gatidis, and B.~Yang.
\newblock Medgan: Medical image translation using gans.
\newblock {\em Computerized medical imaging and graphics}, 79:101684, 2020.

\bibitem{dorjsembe2023conditional}
Z.~Dorjsembe, H.-K. Pao, S.~Odonchimed, and F.~Xiao.
\newblock Conditional diffusion models for semantic 3d medical image synthesis.
\newblock {\em Authorea Preprints}, 2023.

\bibitem{goodfellow2014generative}
I.~Goodfellow, J.~Pouget-Abadie, M.~Mirza, B.~Xu, D.~Warde-Farley, S.~Ozair, A.~Courville, and Y.~Bengio.
\newblock Generative adversarial nets.
\newblock {\em NeurIPS}, 2014.

\bibitem{gretton2012mmd}
A.~Gretton, K.~M. Borgwardt, M.~J. Rasch, B.~Schölkopf, and A.~Smola.
\newblock A kernel two-sample test.
\newblock {\em Journal of Machine Learning Research}, 13:723--773, 2012.

\bibitem{gungor2023adaptive}
A.~G{\"u}ng{\"o}r, S.~U. Dar, {\c{S}}.~{\"O}zt{\"u}rk, Y.~Korkmaz, H.~A. Bedel, G.~Elmas, M.~Ozbey, and T.~{\c{C}}ukur.
\newblock Adaptive diffusion priors for accelerated mri reconstruction.
\newblock {\em MIA}, 88:102872, 2023.

\bibitem{ho2020denoising}
J.~Ho, A.~Jain, and P.~Abbeel.
\newblock Denoising diffusion probabilistic models.
\newblock In {\em NeurIPS}, 2020.

\bibitem{hore2010psnr}
A.~Hore and D.~Ziou.
\newblock Image quality metrics: Psnr vs. ssim.
\newblock In {\em 2010 20th International Conference on Pattern Recognition}, pages 2366--2369, 2010.

\bibitem{hu2021lora}
E.~J. Hu, Y.~Shen, P.~Wallis, Z.~Allen-Zhu, Y.~Li, S.~Wang, L.~Wang, and W.~Chen.
\newblock Lora: Low-rank adaptation of large language models.
\newblock {\em arXiv}, 2021.

\bibitem{jiang2023cola}
L.~Jiang, Y.~Mao, X.~Wang, X.~Chen, and C.~Li.
\newblock Cola-diff: Conditional latent diffusion model for multi-modal mri synthesis.
\newblock In {\em MICCAI}, 2023.

\bibitem{kazeminia2020gans}
S.~Kazeminia, C.~Baur, A.~Kuijper, B.~van Ginneken, N.~Navab, S.~Albarqouni, and A.~Mukhopadhyay.
\newblock Gans for mia.
\newblock {\em Artificial intelligence in medicine}, 109:101938, 2020.

\bibitem{khader2023denoising}
F.~Khader, G.~M{\"u}ller-Franzes, S.~Tayebi~Arasteh, T.~Han, C.~Haarburger, M.~Schulze-Hagen, P.~Schad, S.~Engelhardt, B.~Bae{\ss}ler, S.~Foersch, et~al.
\newblock Denoising diffusion probabilistic models for 3d medical image generation.
\newblock {\em Scientific Reports}, 13(1):7303, 2023.

\bibitem{kim2024adaptive}
J.~Kim and H.~Park.
\newblock Adaptive latent diffusion model for 3d medical image to image translation: Multi-modal magnetic resonance imaging study.
\newblock In {\em WACV}, 2024.

\bibitem{konz2024anatomically}
N.~Konz, Y.~Chen, H.~Dong, and M.~A. Mazurowski.
\newblock Anatomically-controllable medical image generation with segmentation-guided diffusion models.
\newblock {\em arXiv}, 2024.

\bibitem{lyu2022conversion}
Q.~Lyu and G.~Wang.
\newblock Conversion between ct and mri images using diffusion and score-matching models.
\newblock {\em arXiv}, 2022.

\bibitem{ma2024segment}
J.~Ma, Y.~He, F.~Li, L.~Han, C.~You, and B.~Wang.
\newblock Segment anything in medical images.
\newblock {\em Nature Communications}, 15(1):654, 2024.

\bibitem{mehta2021qu}
R.~Mehta and A.~Filos.
\newblock Qu-brats: Miccai brats 2020 challenge on quantifying uncertainty in brain tumor segmentation-analysis of ranking scores and benchmarking results.
\newblock {\em arXiv}, 2021.

\bibitem{69752101}
B.~H. Menze and A.~Jakab.
\newblock The multimodal brain tumor image segmentation benchmark (brats).
\newblock {\em TMI}, 34(10):1993--2024, 2015.

\bibitem{nie2018medical}
D.~Nie, R.~Trullo, J.~Lian, L.~Wang, C.~Petitjean, S.~Ruan, Q.~Wang, and D.~Shen.
\newblock Medical image synthesis with deep convolutional adversarial networks.
\newblock {\em TBE}, 65(12):2720--2730, 2018.

\bibitem{pinaya2022brain}
W.~H. Pinaya, P.-D. Tudosiu, J.~Dafflon, P.~F. Da~Costa, V.~Fernandez, P.~Nachev, S.~Ourselin, and M.~J. Cardoso.
\newblock Brain imaging generation with latent diffusion models.
\newblock In {\em MICCAIW}, 2022.

\bibitem{rahman2023ambiguous}
A.~Rahman, J.~M.~J. Valanarasu, I.~Hacihaliloglu, and V.~M. Patel.
\newblock Ambiguous medical image segmentation using diffusion models.
\newblock In {\em CVPR}, 2023.

\bibitem{rombach2022high}
R.~Rombach, A.~Blattmann, D.~Lorenz, P.~Esser, and B.~Ommer.
\newblock High-resolution image synthesis with latent diffusion models.
\newblock In {\em CVPR}, 2022.

\bibitem{ronneberger2015u}
O.~Ronneberger, P.~Fischer, and T.~Brox.
\newblock U-net: Convolutional networks for biomedical image segmentation.
\newblock In {\em MICCAI}, 2015.

\bibitem{ruiz2023dreambooth}
N.~Ruiz, Y.~Li, V.~Jampani, Y.~Pritch, M.~Rubinstein, and K.~Aberman.
\newblock Dreambooth: Fine tuning text-to-image diffusion models for subject-driven generation.
\newblock In {\em CVPR}, 2023.

\bibitem{skandarani2023gans}
Y.~Skandarani, P.-M. Jodoin, and A.~Lalande.
\newblock Gans for medical image synthesis: An empirical study.
\newblock {\em Journal of Imaging}, 9(3):69, 2023.

\bibitem{sohl2015deep}
J.~Sohl-Dickstein, E.~Weiss, N.~Maheswaranathan, and S.~Ganguli.
\newblock Deep unsupervised learning using nonequilibrium thermodynamics.
\newblock In {\em ICML}, 2015.

\bibitem{ssim}
Z.~Wang, A.~C. Bovik, H.~R. Sheikh, and E.~P. Simoncelli.
\newblock Image quality assessment: from error visibility to structural similarity.
\newblock {\em IEEE transactions on image processing}, 13(4):600--612, 2004.

\bibitem{wang2003msssim}
Z.~Wang, E.~P. Simoncelli, and A.~C. Bovik.
\newblock Multiscale structural similarity for image quality assessment.
\newblock In {\em The Thirty-Seventh Asilomar Conference on Signals, Systems \& Computers, 2003}, volume~2, pages 1398--1402, 2003.

\bibitem{wiatrak2019stabilizing}
M.~Wiatrak, S.~V. Albrecht, and A.~Nystrom.
\newblock Stabilizing generative adversarial networks: A survey.
\newblock {\em arXiv}, 2019.

\bibitem{wolleb2022diffusion}
J.~Wolleb, R.~Sandk{\"u}hler, F.~Bieder, P.~Valmaggia, and P.~C. Cattin.
\newblock Diffusion models for implicit image segmentation ensembles.
\newblock In {\em MIDL}, 2022.

\bibitem{wu2024medsegdiff}
J.~Wu, R.~Fu, H.~Fang, Y.~Zhang, Y.~Yang, H.~Xiong, H.~Liu, and Y.~Xu.
\newblock Medsegdiff: Medical image segmentation with diffusion probabilistic model.
\newblock In {\em MIDL}, 2024.

\bibitem{yurt2022progressively}
M.~Yurt, M.~{\"O}zbey, S.~U. Dar, B.~Tinaz, K.~K. Oguz, and T.~{\c{C}}ukur.
\newblock Progressively volumetrized deep generative models for data-efficient contextual learning of mr image recovery.
\newblock {\em MIA}, 78:102429, 2022.

\bibitem{zhang2023adding}
L.~Zhang, A.~Rao, and M.~Agrawala.
\newblock Adding conditional control to text-to-image diffusion models.
\newblock In {\em ICCV}, 2023.

\bibitem{zhuang2023semantic}
Y.~Zhuang, B.~Hou, T.~S. Mathai, P.~Mukherjee, B.~Kim, and R.~M. Summers.
\newblock Semantic image synthesis for abdominal ct.
\newblock In {\em MICCAI}, 2023.

\end{thebibliography}
